# A Micropolar Peridynamic Theory in Linear Elasticity

S Roy Chowdhury, Md Masiur Rahaman, Debasish Roy[*] and Narayan Sundaram

Department of Civil Engineering, Indian Institute of Science, Bangalore 560012, India

(Corresponding author; email: royd@civil.iisc.ernet.in)

**Abstract**

*A state-based micropolar peridynamic theory for linear elastic solids is proposed. The main motivation is to introduce additional micro-rotational degrees of freedom to each material point and thus naturally bring in the physically relevant material length scale parameters into peridynamics. Non-ordinary type modeling via constitutive correspondence is adopted here to define the micropolar peridynamic material. Along with a general three dimensional model, homogenized one dimensional Timoshenko type beam models for both the proposed micropolar and the standard non-polar peridynamic variants are derived. The efficacy of the proposed models in analyzing continua with length scale effects is established via numerical simulations of a few beam and plane-stress problems.*

**Keywords**: *micropolar peridynamics; length scales; constitutive correspondence; Timoshenko-type beams; plane stress problems*

## 1. Introduction

Classical continuum mechanics assumes a continuous distribution of matter throughout the body and establishes the equations of motion considering only the local action. Both long range effects of loads and those of inter-molecular interactions are ignored in this theory. Such approximations limit the applicability of the classical theory to macro-scale phenomena where the characteristic length scale of the loading is much larger than the intrinsic material length scales. However, when these length scales are comparable, the microstructural effects could become significant and predictions of the classical theory depart considerably from experimental results. Substantial discrepancies are observed, for instance, in problems involving high stress gradients at notch and crack tips, short wavelength dynamic excitations, the behavior of granular solids, porous materials, modern-day engineering nano-structures etc (Eringen, 1976).

In order to circumvent such limitations of the classical continuum theory of elasticity, an early attempt was made by Voigt (1887), who postulated the existence of a couple-traction along with the usual force-traction responsible for the force transfer across boundaries/interfaces or from one part of the body to another. Later Cosserat and Cosserat (1909) developed a mathematical model based on couple stresses leading to a description of the stress fields via asymmetric tensors as opposed to the symmetric Cauchy stress fields in the classical theory. From a kinematical perspective, this theory enables non-local interactions via the incorporation of rotational degrees of freedom, along with the classically employed translational ones, for the material points and this allows an infinitesimal volume element about a material point to rotate independently of the translational motion. This idea, as formalized in the micropolar theory (Eringen, 1999), assumes the material micro-rotation to be independent of the continuum macro-rotation (e.g. the curl of the displacement field). Such a microstructure-motivated description of deformation provided for the inclusion of length scale parameters in the constitutive equations which were otherwise absent. The development of a structured generalized continuum theory only took place several decades later. Among numerous such contributions, we cite (Eringen and Suhubi, 1964, Nowacki, 1970, Kafadar and Eringen, 1971, Eringen and Kafadar, 1976, Eringen, 1999) and the references therein.

Gradient type non-local formulations (Mindlin, 1965, Mindlin and Eshel, 1968) are yet another approach to a generalized continuum theory where, in lieu of micro-rotations, several higher order gradients of the strain tensor are assumed to contribute to the internal work thereby bringing in the length scale effect. As a consequence, such formulations introduce different higher order generalized stresses conjugate to the gradients of strain.

Apart from the limitation related to scale independence, classical continuum mechanics is not applicable to several other problems of fundamental interest in solid mechanics, viz. those including discontinuities such as existing cracks or spontaneously emerging and propagating micro-cracks and voids, dislocation dynamics in polycrystalline solids etc. The inability to track and evolve a discontinuous field may be traced back to the kinematical requirement of a smooth, diffeomorphism-type deformation field, typically yielding equations of motion in the form of hyperbolic partial differential equations (PDEs). Therefore computational methods for solving such problems using the classical theory either require a redefinition of the object manifold so that discontinuities lie on the boundary or some special treatment to define the spatial derivatives of the field variables on a cracked surface (Bittencourt et al., 1996, Belytschko and Black, 1999, Areias and Belytschko, 2005).

More recently Silling (2000) introduced a continuum theory, the Peridynamics (PD), which is capable of addressing problems involving discontinuities and/or long range forces. One of the main features of this theory is the representation of the equations of motion through integro-differential equations instead of PDEs. This relaxes, to a significant extent, the smoothness requirement of the deformation field and even allows for discontinuities as long as the Riemann integrability of the spatial integrals is ensured. These equations are based on a model of internal forces that the material points exert on each other over finite distances. The initial model, the *bond-based* PD, treats the internal forces as a network of interacting pairs like springs. The maximum distance through which a material particle interacts with its neighbors via spring like interactions is denoted as the *horizon*. Such pair-wise forces however lead to an oversimplification of the model and in particular results in an effective Poisson's ratio of $1/4$ for linear isotropic elastic materials. This limitation has been overcome through a more general model, the *state-based* PD (Silling et al., 2007). According to this theory, particles interact via bond forces that are no longer governed by a central potential independent of the behavior of other bonds; instead they are determined by the collective deformations of the bonds within the horizon of a material particle. This version of the PD theory is applicable over the entire permissible range of Poisson's ratio. Even though the PD has many attractive features, the scarcity of strictly PD-based material constitutive models tends to limit its applicability. This difficulty may however be bypassed using a constitutive correspondence framework (Silling et al., 2007), which enables the use of classical material models in a PD formulation.

In the present work, a novel proposal for a PD approach incorporating micropolar elasticity is set forth. A set of state-based equations of motion is derived for the micropolar continuum and the constitutive correspondence utilized to define the associated material model. Incorporation of additional physical information via the material length-scale parameters has been a primary motivation in the current development. Such an enhancement of the model is expected to emulate more closely the physical behavior of structures like nano-beams, nano-sheets, fracture characteristics of thin films, concrete structures etc. In this context, an earlier work by Gerstle *et al*. (2007) on bond-based micropolar PD should be mentioned, which, whilst eliminating the issue of fixed Poisson's ratio, does not offer a ready framework to incorporate the rich repertoire of classical material models. The last work also has additional limitations in imposing the incompressibility constraint, often employed in a wide range of models including those involving plastic deformation in metals (Silling et al., 2007). Along with a general three dimensional model, a one dimensional micropolar PD model for a Timoshenko type beam is also derived in this work through an appropriate dimensional descent. For the purpose of comparison, a similar beam model based on the standard non-polar PD is derived. Effects of the length scale parameters on the static deformation characteristics of a beam under different boundary conditions are numerically assessed confirming the superiority of the (new) micropolar model over the non-polar one. A two dimensional planar problem of a plate with a hole under tensile loading also holds out similar observations. The theoretical development in this article is, however, limited to linear isotropic elastic deformations only.

The rest of the paper is organized as follows. Section 2 briefly describes the state based PD theory and also gives a short account of linear elastic micropolar theory. While Section 3 reports on a systematic derivation of a general 3D micropolar PD theory, the one dimensional adaptations of the theory are laid out in Sections 4 and 5. This is followed by numerical illustrations and a few concluding remarks in Sections 6 and 7 respectively.

## 2. State Based PD and Micropolar Elasticity

For completeness, a concise description of the state based PD theory along with the constitutive correspondence is given in this section. Equations of motion in the micropolar theory and the linear elastic material model are also briefly reviewed.

### *2.1 State based PD*

Following the approach in (Silling et al., 2007), a brief account of the state-based PD theory is presented below. PD is a non-local continuum theory that describes the dynamics of a body occupying a region $\mathcal{B}_o \subset \mathbb{R}^3$ in its reference configuration and $\mathcal{B}_t \subset \mathbb{R}^3$ in the current configuration. A schematic of the body is shown in Figure 1. The *bond* vector $\xi$ between a material point $X \in \mathcal{B}_o$ and its neighbor $X' \in \mathcal{B}_o$, defined as $\xi = X' - X$, gets deformed under the deformation map $\chi : \mathcal{B}_o \to \mathcal{B}_t$. The deformed bond is given by the deformation vector state $\underline{\mathbf{Y}}$ (refer to (Silling et al., 2007) for a precise definition of states)

$$\underline{\mathbf{Y}}[X]\langle \xi \rangle = y' - y = \chi(X') - \chi(X) \tag{1}$$

The family of bonds to be considered for a point $X$ is given by its *horizon* $\mathcal{H}$ defined as $\mathcal{H}(X) = \{\xi \in \mathbb{R}^3 \mid (\xi + X) \in \mathcal{B}_o, |\xi| < \delta\}$, where $\delta > 0$ is the radius of the horizon.

The state based PD equations of motion are of the following integro-differential form.

$$\rho(X)\ddot{y}(X, t) = \int_{\mathcal{H}(X)} \{\underline{\mathbf{T}}[X, t]\langle \xi \rangle - \underline{\mathbf{T}}[X+\xi, t]\langle -\xi \rangle\} dV_{X'} + b(X, t), \tag{2}$$

where $\rho, \underline{\mathbf{T}}, b$ are the mass density, long range internal force vector state and externally applied body force density respectively. Superimposed dots indicate material derivatives with respect to time. This equation has been shown in (Silling et al., 2007) to satisfy the linear momentum balance. In the standard non-polar PD theory, conservation of angular momentum is ensured by imposing the following restriction on the constitutive relation.

$$\int_{\mathcal{H}(X)} \underline{\mathbf{T}}[X, t]\langle \xi \rangle \times \underline{\mathbf{Y}}[X, t]\langle \xi \rangle dV_{X'} = 0, \quad \forall X \in \mathcal{B}_o \tag{3}$$

Silling *et al.* (2007) have proposed a constitutive correspondence in order to incorporate the classical material models within the PD framework. This is obtained via a non-local definition of the deformation gradient in terms of the deformation state and then equating the PD-based internal energy to the classical one for the same deformation. The correspondence relations are as shown below.

$$\bar{F}(\underline{Y}) = \left[ \int_{\mathcal{H}} \omega(|\xi|)(\underline{Y}\langle\xi\rangle \otimes \xi) dV_{X'} \right] \bar{K}^{-1} \tag{4}$$

$$\bar{K} = \int_{\mathcal{H}} \omega(|\xi|)(\xi \otimes \xi) dV_{X'} \tag{5}$$

$$\underline{T}\langle\xi\rangle = \omega(|\xi|) \bar{P} \bar{K}^{-1} \xi \tag{6}$$

Here $\bar{F} \in \mathbb{R}^3 \times \mathbb{R}^3$ is the non-local deformation gradient tensor, $\bar{K} \in \mathbb{R}^3 \times \mathbb{R}^3$ a non-local shape tensor, $\omega$ a non-negative scalar influence function ($\omega(|\xi|) > 0, \forall \xi \in \mathcal{H}$) and $\bar{P} = \tilde{P}(\bar{F})$ the first Piola-Kirchhoff stress obtained from the classical constitutive relation for $\tilde{P}$ written in terms of the non-local deformation gradient $\bar{F}$. It has also been shown (Silling et al., 2007) that such correspondence is consistent with Eq. (3) and hence satisfies the conservation of angular momentum.

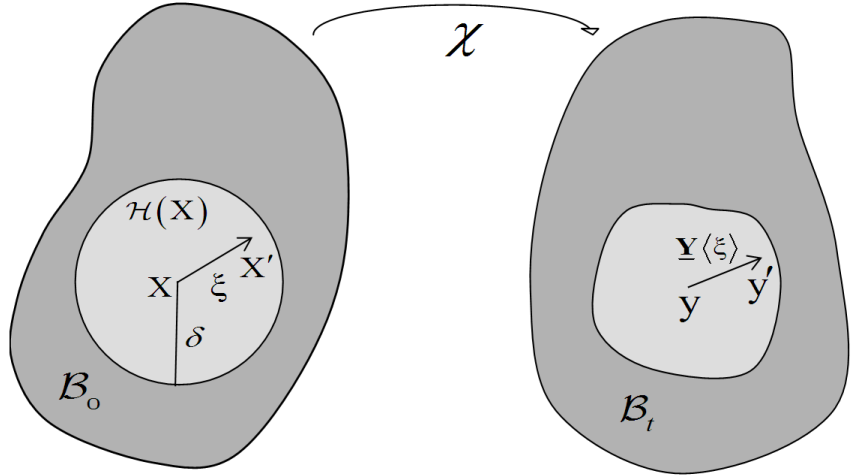

Figure 1. Schematic PD body in the reference and current configurations

*2.2 Micropolar Elasticity*

Linear elastic micropolar theory for an isotropic material is briefly reviewed here. For a more detailed exposition, the reader is referred to Eringen (1999). The micropolar kinematical relations are given through

$$\gamma_{ji} = u_{i,j} - \varepsilon_{kji}\theta_k , \quad \kappa_{ji} = \theta_{i,j} , \quad i, j, k = 1, 2, 3 \tag{7}$$

where $\gamma$ is the infinitesimal micropolar strain tensor and $\kappa$ the infinitesimal wryness tensor. $u$ and $\theta$ respectively denote the displacement and micro-rotation vectors and $\varepsilon$ is the Levi-Civita symbol or the alternate tensor. $u_{i,j}$ denotes the partial derivative of $u_i$ with respect to the spatial coordinate $X_j$. Einstein summation convention holds for the expressions having repeated indices.

Micropolar equations of motions are obtained from the conservation of linear and angular momenta and they are written in the following form.

$$\sigma_{ji,j} + b_i = \rho \ddot{u}_i \quad , \quad \mu_{ji,j} + \varepsilon_{ijk}\sigma_{jk} + l_i = J\ddot{\theta}_i \, , \tag{8}$$

where $\sigma$ is asymmetric stress tensor and $\mu$ the couple stress tensor. $b, l, J$ respectively refer to the external body force density, body couple density and micro-inertia. In linear elastic micropolar theory, these equations of motion are supplemented with the following constitutive relations.

$$\sigma_{ji} = \lambda \gamma_{kk} \delta_{ji} + (\tilde{\mu} + \eta)\gamma_{ji} + \tilde{\mu}\gamma_{ij} \quad , \quad \mu_{ji} = \alpha \kappa_{kk} \delta_{ji} + \beta \kappa_{ji} + \gamma \kappa_{ij} \, . \tag{9}$$

Here $\lambda$ and $\tilde{\mu}$ are the Lamé constants, and $\eta, \alpha, \beta, \gamma$ are micropolar material constants. More familiar constants like the shear modulus $G$, Young's modulus $E$ and Poisson's ratio $\nu$ are defined in terms of these parameters as follows.

$$G = \tilde{\mu} + \frac{\eta}{2} \, , \quad E = \frac{G(3\lambda + 2G)}{\lambda + G} \, , \quad \nu = \frac{\lambda}{2\lambda + 2G} \, . \tag{10}$$

## 3. A Micropolar PD Theory

In developing a micropolar PD theory, we assume the existence of a non-local internal moment state $\underline{\mu}$ that is generated along with the internal force state $\underline{T}$ as a response to the externally applied body force and body couple.

*3.1 Kinematics*

Following the discussion on micropolar elasticity in Section 2.2, the micro-rotation $\theta$ is introduced as additional degrees of freedom within the PD continuum in order to allow for rigid-type rotation of each material point. For a kinematical description of the PD body, we need to define a few new vector states as listed below.

Relative displacement state, $\underline{U}[X]\langle\xi\rangle = u' - u = (\chi(X') - X') - (\chi(X) - X)$ (11)

Relative rotation state, $\underline{\Theta}[X]\langle\xi\rangle = \theta' - \theta = \theta(X') - \theta(X)$ (12)

Averaged rotation state, $\underline{\hat{\Theta}}[X]\langle\xi\rangle = \frac{1}{2}(\theta' + \theta)$ (13)

*3.2 Equations of motion*

For the micropolar PD body, we postulate that the following additional equations, along with the equations of motion given in Eq. (2), must be satisfied.

$$J(X)\ddot{\theta}(X,t) = \int_{\mathcal{H}(X)} \{\underline{\mu}\langle\xi\rangle - \underline{\mu}'\langle-\xi\rangle\} dV_{X'} + \frac{1}{2}\int_{\mathcal{H}(X)} \underline{Y}\langle\xi\rangle \times \{\underline{T}\langle\xi\rangle - \underline{T}'\langle-\xi\rangle\} dV_{X'} + l(X,t) , \quad (14)$$

where $\underline{\mu}'\langle-\xi\rangle = \underline{\mu}[X+\xi, t]\langle-\xi\rangle$ and a similar definition holds for $\underline{T}'$. l is the externally applied body couple density.

In order to be physically admissible, it is necessary that the states $\underline{\mu}$ and $\underline{T}$ satisfy the balance of linear and angular momenta for any bounded body $\mathcal{B}$. While the former (i.e. the balance of linear momentum) is known to be satisfied by the specific form of the equations of motion adopted in Eq. (2), the latter requires the satisfaction of the following identity.

$$\int_{\mathcal{B}} y \times \rho(X)\ddot{y}(X,t) dV_X + \int_{\mathcal{B}} J(X)\ddot{\theta}(X,t) dV_X = \int_{\mathcal{B}} y \times b(X,t) dV_X + \int_{\mathcal{B}} l(X,t) dV_X \quad (15)$$

**Proposition 3.1** *Let the bounded body $\mathcal{B}$ be subjected to body force and body couple density* b *and* l *respectively. Let the internal force and moment vector states be $\underline{\mu}$ and $\underline{T}$ respectively. Then if equation (14) holds in $\mathcal{B}$, balance of angular momentum will be satisfied, i.e., Eq. (15) will hold.*

*Proof:* Substituting Eqs. (2) and (14) in the left hand side of Eq. (15), one obtains the following equation: (with notational abuse, e.g., $\underline{T}$ is used in place of $\underline{T}\langle\xi\rangle$ and similarly for the other states).

$$\int_{\mathcal{B}} y \times \rho(X)\ddot{y}(X,t) dV_X + \int_{\mathcal{B}} J(X)\ddot{\theta}(X,t) dV_X = \int_{\mathcal{B}}\int_{\mathcal{H}} y \times \{\underline{T}-\underline{T}'\} dV_{X'} dV_X + \int_{\mathcal{B}} y \times b \, dV_X$$
$$+ \int_{\mathcal{B}}\int_{\mathcal{H}} \{\underline{\mu}-\underline{\mu}'\} dV_{X'} dV_X + \frac{1}{2}\int_{\mathcal{B}}\int_{\mathcal{H}} (y'-y) \times \{\underline{T}-\underline{T}'\} dV_{X'} dV_X + \int_{\mathcal{B}} l \, dV_X \quad (16)$$

Since no interaction exists beyond the horizon ($\mathcal{H}$), all the inner integrals can be extended over the entire body ($\mathcal{B}$) without affecting the result. A change of variable $X \leftrightarrow X'$ and the subsequent application of Fubini's theorem lead to the following simplifications:

$$\int_{\mathcal{B}}\int_{\mathcal{H}} \{\underline{\mu}-\underline{\mu}'\} dV_{X'} dV_X = \int_{\mathcal{B}}\int_{\mathcal{B}} \{\underline{\mu}-\underline{\mu}'\} dV_{X'} dV_X = \int_{\mathcal{B}}\int_{\mathcal{B}} \{\underline{\mu}'-\underline{\mu}\} dV_{X'} dV_X = 0 \quad (17)$$

$$\frac{1}{2}\int_{\mathcal{B}}\int_{\mathcal{H}} (y'-y) \times \{\underline{T}-\underline{T}'\} dV_{X'} dV_X = \frac{1}{2}\int_{\mathcal{B}}\int_{\mathcal{B}} (y'-y) \times \{\underline{T}-\underline{T}'\} dV_{X'} dV_X$$
$$= \frac{1}{2}\int_{\mathcal{B}}\int_{\mathcal{B}} y' \times \{\underline{T}-\underline{T}'\} dV_{X'} dV_X - \frac{1}{2}\int_{\mathcal{B}}\int_{\mathcal{B}} y \times \{\underline{T}-\underline{T}'\} dV_{X'} dV_X$$
$$= \frac{1}{2}\int_{\mathcal{B}}\int_{\mathcal{B}} y \times \{\underline{T}'-\underline{T}\} dV_{X'} dV_X - \frac{1}{2}\int_{\mathcal{B}}\int_{\mathcal{B}} y \times \{\underline{T}-\underline{T}'\} dV_{X'} dV_X$$
$$= -\int_{\mathcal{B}}\int_{\mathcal{B}} y \times \{\underline{T}-\underline{T}'\} dV_{X'} dV_X \quad (18)$$

Using Eqs. (17) and (18), Eq. (16) gets simplified to Eq. (15) and thus establishes the balance of angular momentum. □

It is to be noted that for problems with small elastic deformation, the second term of right hand side of Eq.(14) could be approximated by $\frac{1}{2}\int_{\mathcal{H}(X)} \xi \times \{\underline{T}\langle\xi\rangle - \underline{T}'\langle-\xi\rangle\} dV_{X'}$ .

### 3.3 Energy balance and constitutive relations

We will focus here only on the mechanical energy balance without considering any heat source or flux. Taking scalar products of both sides of Eq. (2) with the velocity $\dot{y}$ and Eq. (14) with $\dot{\theta}$ and their subsequent addition and integration over a finite sub-region $\mathcal{P} \subset \mathcal{B}$ result in the following.

$$\frac{d}{dt}\int_{\mathcal{P}} \frac{\rho \dot{y} \cdot \dot{y}}{2} dV_X + \frac{d}{dt}\int_{\mathcal{P}} \frac{J\dot{\theta} \cdot \dot{\theta}}{2} dV_X = \int_{\mathcal{P}}\int_{\mathcal{B}} \{\underline{T} - \underline{T}'\} \cdot \dot{y}\, dV_{X'} dV_X + \frac{1}{2}\int_{\mathcal{P}}\int_{\mathcal{B}} (y' - y) \times \{\underline{T} - \underline{T}'\} \cdot \dot{\theta}\, dV_{X'} dV_X$$
$$+ \int_{\mathcal{P}}\int_{\mathcal{B}} \{\underline{\mu} - \underline{\mu}'\} \cdot \dot{\theta}\, dV_{X'} dV_X + \int_{\mathcal{P}} b \cdot \dot{y}\, dV_X + \int_{\mathcal{P}} l \cdot \dot{\theta}\, dV_X \qquad (19)$$

Note that all the inner integrals in the expression above have been extended to the whole body since there is no interaction beyond the horizon, as indicated earlier.

The following identities (Eqs. (20)-(22)) are needed for further simplification.

$$\{\underline{T}\langle\xi\rangle - \underline{T}'\langle\xi\rangle\} \cdot \dot{y} = (\underline{T}\langle\xi\rangle \cdot \dot{y}' - \underline{T}'\langle\xi\rangle \cdot \dot{y}) - \underline{T}\langle\xi\rangle \cdot (\dot{y}' - \dot{y}) \qquad (20)$$

$$\frac{1}{2}(y' - y) \times \{\underline{T}\langle\xi\rangle - \underline{T}'\langle\xi\rangle\} \cdot \dot{\theta} = \left(\underline{T}\langle\xi\rangle \cdot \frac{\dot{\theta}'}{2} \times (y - y') - \underline{T}'\langle\xi\rangle \cdot \frac{\dot{\theta}}{2} \times (y' - y)\right) + \underline{T}\langle\xi\rangle \cdot \frac{\dot{\theta} + \dot{\theta}'}{2} \times (y' - y) \qquad (21)$$

$$\{\underline{\mu}\langle\xi\rangle - \underline{\mu}'\langle\xi\rangle\} \cdot \dot{\theta} = (\underline{\mu}\langle\xi\rangle \cdot \dot{\theta}' - \underline{\mu}'\langle\xi\rangle \cdot \dot{\theta}) - \underline{\mu}\langle\xi\rangle \cdot (\dot{\theta}' - \dot{\theta}) \qquad (22)$$

Using the above identities, each of first three terms on the right hand side of Eq. (19) could be split into two integrals as illustrated below.

$$\int_{\mathcal{P}}\int_{\mathcal{B}} \{\underline{T} - \underline{T}'\} \cdot \dot{y}\, dV_{X'} dV_X = \int_{\mathcal{P}}\int_{\mathcal{B}} (\underline{T} \cdot \dot{y}' - \underline{T}' \cdot \dot{y}) dV_{X'} dV_X - \int_{\mathcal{P}}\int_{\mathcal{B}} \underline{T} \cdot (\dot{y}' - \dot{y}) dV_{X'} dV_X$$
$$= \int_{\mathcal{P}}\int_{\mathcal{B}\setminus\mathcal{P}} (\underline{T} \cdot \dot{y}' - \underline{T}' \cdot \dot{y}) dV_{X'} dV_X - \int_{\mathcal{P}}\int_{\mathcal{B}} \underline{T} \cdot (\dot{y}' - \dot{y}) dV_{X'} dV_X, \qquad (23)$$

where the antisymmetry of the integrand in first integral is used to obtain the last step. Similarly,

$$\frac{1}{2}\int_{\mathcal{P}}\int_{\mathcal{B}} (y' - y) \times \{\underline{T} - \underline{T}'\} \cdot \dot{\theta}\, dV_{X'} dV_X$$
$$= \int_{\mathcal{P}}\int_{\mathcal{B}\setminus\mathcal{P}} \left(\underline{T} \cdot \frac{\dot{\theta}'}{2} \times (y - y') - \underline{T}' \cdot \frac{\dot{\theta}}{2} \times (y' - y)\right) dV_{X'} dV_X + \int_{\mathcal{P}}\int_{\mathcal{B}} \underline{T} \cdot \frac{\dot{\theta} + \dot{\theta}'}{2} \times (y' - y) dV_{X'} dV_X \qquad (24)$$

and

$$\int_{\mathcal{P}}\int_{\mathcal{B}} \{\underline{\mu} - \underline{\mu}'\} \cdot \dot{\theta}\, dV_{X'} dV_X = \int_{\mathcal{P}}\int_{\mathcal{B}\setminus\mathcal{P}} (\underline{\mu} \cdot \dot{\theta}' - \underline{\mu}' \cdot \dot{\theta}) dV_{X'} dV_X - \int_{\mathcal{P}}\int_{\mathcal{B}} \underline{\mu} \cdot (\dot{\theta}' - \dot{\theta}) dV_{X'} dV_X. \qquad (25)$$

Using Eqs. (23)-(25), Eq. (19) can be rewritten in the following form, representing a power balance.

$$\dot{\mathcal{K}}(\mathcal{P}) + \mathcal{W}_{abs}(\mathcal{P}) = \mathcal{W}_{sup}(\mathcal{P}), \qquad (26)$$

where $\mathcal{K}(\mathcal{P}) = \int_{\mathcal{P}} \frac{\rho \dot{\mathbf{y}} \cdot \dot{\mathbf{y}}}{2} dV_X + \int_{\mathcal{P}} \frac{J\dot{\theta} \cdot \dot{\theta}}{2} dV_X$ is the kinetic energy in $\mathcal{P}$. The absorbed power in $\mathcal{P}$ and the supplied power to $\mathcal{P}$ are respectively given by

$$\mathcal{W}_{abs}(\mathcal{P}) = \int_{\mathcal{P}} \int_{\mathcal{B}} \underline{\mathbf{T}} \cdot \left\{ (\dot{\mathbf{y}}' - \dot{\mathbf{y}}) - \frac{\dot{\theta} + \dot{\theta}'}{2} \times (\mathbf{y}' - \mathbf{y}) \right\} dV_{X'} dV_X + \int_{\mathcal{P}} \int_{\mathcal{B}} \underline{\boldsymbol{\mu}} \cdot (\dot{\theta}' - \dot{\theta}) dV_{X'} dV_X \tag{27}$$

$$\mathcal{W}_{sup}(\mathcal{P}) = \int_{\mathcal{P}} \int_{\mathcal{B}\setminus\mathcal{P}} \left( \underline{\mathbf{T}} \cdot \left( \dot{\mathbf{y}}' + \frac{\dot{\theta}'}{2} \times (\mathbf{y} - \mathbf{y}') \right) - \underline{\mathbf{T}}' \cdot \left( \dot{\mathbf{y}} + \frac{\dot{\theta}}{2} \times (\mathbf{y}' - \mathbf{y}) \right) \right) dV_{X'} dV_X$$

$$+ \int_{\mathcal{P}} \int_{\mathcal{B}\setminus\mathcal{P}} \left( \underline{\boldsymbol{\mu}} \cdot \dot{\theta}' - \underline{\boldsymbol{\mu}}' \cdot \dot{\theta} \right) dV_{X'} dV_X + \int_{\mathcal{P}} \mathbf{b} \cdot \dot{\mathbf{y}} \, dV_X + \int_{\mathcal{P}} \mathbf{l} \cdot \dot{\theta} \, dV_X \tag{28}$$

Since no other source of energy is considered, the absorbed power relates to the internal energy $\mathcal{E}(\mathcal{P})$ as $\dot{\mathcal{E}}(\mathcal{P}) = \mathcal{W}_{abs}(\mathcal{P})$. Also the form of integrals in (27) shows the additive character of the internal energy (i.e., $\mathcal{E}(\mathcal{P}_1) + \mathcal{E}(\mathcal{P}_2) = \mathcal{E}(\mathcal{P}_1 + \mathcal{P}_2)$) thus ensuring the existence of an internal energy density, $e$ which therefore has the following rate form.

$$\dot{e} = \int_{\mathcal{B}} \underline{\mathbf{T}} \cdot \left\{ (\dot{\mathbf{y}}' - \dot{\mathbf{y}}) - \frac{\dot{\theta} + \dot{\theta}'}{2} \times (\mathbf{y}' - \mathbf{y}) \right\} dV_{X'} + \int_{\mathcal{B}} \underline{\boldsymbol{\mu}} \cdot (\dot{\theta}' - \dot{\theta}) dV_{X'} \tag{29}$$

For small deformation, we introduce an approximation to the above expression leading to

$$\dot{e} = \int_{\mathcal{B}} \underline{\mathbf{T}} \cdot \left\{ (\dot{\mathbf{y}}' - \dot{\mathbf{y}}) - \frac{\dot{\theta} + \dot{\theta}'}{2} \times (\mathbf{X}' - \mathbf{X}) \right\} dV_{X'} + \int_{\mathcal{B}} \underline{\boldsymbol{\mu}} \cdot (\dot{\theta}' - \dot{\theta}) dV_{X'} . \tag{30}$$

Using the definition of inner product of states given in (Silling et al., 2007), the expression above is rewritten as follows.

$$\dot{e} = \underline{\mathbf{T}} \bullet \left( \underline{\dot{\mathbf{Y}}} - \dot{\hat{\Theta}} \times \underline{\mathbf{X}} \right) + \underline{\boldsymbol{\mu}} \bullet \underline{\dot{\Theta}} = \underline{\mathbf{T}} \bullet \left( \underline{\dot{\mathbf{U}}} - \dot{\hat{\Theta}} \times \underline{\mathbf{X}} \right) + \underline{\boldsymbol{\mu}} \bullet \underline{\dot{\Theta}} = \underline{\mathbf{T}} \bullet \underline{\dot{\mathbf{U}}}_{\hat{\Theta}} + \underline{\boldsymbol{\mu}} \bullet \underline{\dot{\Theta}}, \tag{31}$$

where $\underline{\mathbf{U}}_{\hat{\Theta}} = \underline{\mathbf{U}} - \hat{\Theta} \times \underline{\mathbf{X}}$ is a composite state and its action on a bond $\xi$ is given as $\underline{\mathbf{U}}_{\hat{\Theta}} \langle \xi \rangle = (\mathbf{u}' - \mathbf{u}) - \frac{\theta' + \theta}{2} \times (\mathbf{X}' - \mathbf{X})$.

Assuming the energy density functional $e$ to depend on $\underline{\mathbf{Y}}_{\hat{\Theta}}$ and $\underline{\Theta}$, the rate of $e$ would be given by the following expression.

$$\dot{e}\left(\underline{\mathbf{U}}_{\hat{\Theta}}, \underline{\Theta}\right) = e_{\underline{\mathbf{U}}_{\hat{\Theta}}} \bullet \underline{\dot{\mathbf{U}}}_{\hat{\Theta}} + e_{\underline{\Theta}} \bullet \underline{\dot{\Theta}}, \tag{32}$$

where $e_{\underline{\mathbf{U}}_{\hat{\Theta}}}$ and $e_{\underline{\Theta}}$ are the Fréchet derivatives (Silling et al., 2007) of $e$ with respect to $\underline{\mathbf{U}}_{\hat{\Theta}}$ and $\underline{\Theta}$ respectively. Comparison of (31) with (32) leads to the following constitutive relations (which are consistent with the second law of thermodynamics and also obtainable by the Coleman-Noll procedure (Tadmor et al., 2012)).

$$\underline{\mathbf{T}} = e_{\underline{\mathbf{U}}_{\hat{\Theta}}} \quad \text{and} \quad \underline{\boldsymbol{\mu}} = e_{\underline{\Theta}} \tag{33}$$

### 3.3.1 Constitutive correspondence

In order to make use of the micropolar material model described in Eq. (9), the constitutive correspondence route is adopted and this calls for definitions of non-local strain and wryness tensors. Note that, for a continuously differentiable deformation field on $\mathcal{B}$, the following holds.

$$\underline{\mathbf{U}}_{\hat{\Theta}}\langle\xi\rangle = \left(\mathbf{u}(X+\xi) - \mathbf{u}(X)\right) - \frac{\theta(X+\xi) + \theta(X)}{2} \times \left((X+\xi) - X\right) = (\nabla \mathbf{u} + \varepsilon \cdot \theta)\xi + O\left(|\xi|^2\right) = \gamma\xi + O\left(|\xi|^2\right) \tag{34}$$

$$\underline{\Theta}\langle\xi\rangle = \theta(X+\xi) - \theta(X) = (\nabla\theta)\xi + O\left(|\xi|^2\right) = \kappa\xi + O\left(|\xi|^2\right), \tag{35}$$

where $\nabla$ is the gradient operator, $\varepsilon$ is the third order alternate tensor, and $\gamma$ and $\kappa$ are respectively the infinitesimal micropolar strain and wryness tensors. Analogous to Eq. (4), following are the non-local approximations to $\gamma$ and $\kappa$.

$$\bar{\gamma}(\underline{\mathbf{U}}_{\hat{\Theta}}) = \left[\int_{\mathcal{H}} \omega(|\xi|)\left(\underline{\mathbf{U}}_{\hat{\Theta}}\langle\xi\rangle \otimes \xi\right) dV_{X'}\right]\bar{K}^{-1}, \quad \bar{\kappa}(\underline{\Theta}) = \left[\int_{\mathcal{H}} \omega(|\xi|)\left(\underline{\Theta}\langle\xi\rangle \otimes \xi\right) dV_{X'}\right]\bar{K}^{-1} \tag{36}$$

Let $w(\bar{\gamma},\bar{\kappa})$ be the internal energy density (per unit volume of the micropolar continuum) written in terms of the non-local kinematic quantities. Then equating $e(\underline{\mathbf{U}}_{\hat{\Theta}},\underline{\Theta})$ with $w(\bar{\gamma},\bar{\kappa})$, the PD model for linear elastic micropolar materials is established. Following the computations below, explicit forms of the desired PD material model could be obtained.

$$\dot{e}(\underline{\mathbf{U}}_{\hat{\Theta}},\underline{\Theta}) = \dot{w}(\bar{\gamma},\bar{\kappa}) = \sigma : \dot{\bar{\gamma}} + \mu : \dot{\bar{\kappa}}$$

$$= \sigma : \left[\int_{\mathcal{H}} \omega(|\xi|)\left(\underline{\dot{\mathbf{U}}}_{\hat{\Theta}}\langle\xi\rangle \otimes \xi\right) dV_{X'}\right]\bar{K}^{-1} + \mu : \left[\int_{\mathcal{H}} \omega(|\xi|)\left(\underline{\dot{\Theta}}\langle\xi\rangle \otimes \xi\right) dV_{X'}\right]\bar{K}^{-1}$$

$$= \int_{\mathcal{H}} \omega(|\xi|)\sigma : \left(\underline{\dot{\mathbf{U}}}_{\hat{\Theta}}\langle\xi\rangle \otimes \xi\right)\bar{K}^{-1} dV_{X'} + \int_{\mathcal{H}} \omega(|\xi|)\mu : \left(\underline{\dot{\Theta}}\langle\xi\rangle \otimes \xi\right)\bar{K}^{-1} dV_{X'}$$

$$= \int_{\mathcal{H}} \omega(|\xi|)\sigma\bar{K}^{-1}\xi \cdot \underline{\dot{\mathbf{U}}}_{\hat{\Theta}}\langle\xi\rangle dV_{X'} + \int_{\mathcal{H}} \omega(|\xi|)\mu\bar{K}^{-1}\xi \cdot \underline{\dot{\Theta}}\langle\xi\rangle dV_{X'} \tag{37}$$

Note that the symmetry of shape tensor $\bar{K}$ is utilized in writing the last expression. Comparing (37) with (31), the following constitutive relations are obtained.

$$\underline{\mathbf{T}}\langle\xi\rangle = \omega(|\xi|)\sigma\bar{K}^{-1}\xi \quad \text{and} \quad \underline{\boldsymbol{\mu}}\langle\xi\rangle = \omega(|\xi|)\mu\bar{K}^{-1}\xi, \tag{38}$$

where $\sigma = \lambda \operatorname{tr}(\bar{\gamma})I + (\tilde{\mu}+\eta)\bar{\gamma}^T + \tilde{\mu}\bar{\gamma}$ and $\mu = \alpha \operatorname{tr}(\kappa)I + \beta\kappa + \gamma\kappa^T$ (39)

$\operatorname{tr}(\cdot)$ is the trace operator, $(\cdot)^T$ the transpose operator and $I$ the second order identity tensor.

The equations of motion Eqs. (2) and (14) along with the constitutive model (38) complete the description of linear elastic micropolar PD theory.

**4. 1D Micropolar PD Beam**

Since the discovery of carbon nano-tube in the early 1990s, an intensive research effort has been directed to the analysis of such nano structures. Successful attempts have been made in developing non-local continuum formulations to analyze static and vibration characteristics of nano-rods, nano-beams, nano-films etc. Being a non-local model incorporating material length scale parameters, the proposed micropolar PD model provides a suitable framework for such problems. In the specific context of analyzing nano-rods and nano-beams, a one dimensional version of the proposed model would be of considerable interest, enabling one to obtain results with lesser computational effort as compared to the full-blown 3D model.

A Timoshenko type beam model, set in the micropolar PD framework, is presented here.

*4.1 Geometry and assumptions*

Figure 2 shows a schematic of the beam geometry. A Cartesian coordinate frame is considered with its origin located at one end of the beam. The $x$ axis is chosen along the length and lies on the plane containing neutral axes of the cross sections, $y$ is along the neutral axis of the cross section containing the origin and $z$ is along the thickness. The beam cross section is assumed to be symmetric about the $z$ axis. Our proposal for the PD micropolar beam is based on a few assumptions; viz. the cross-sectional dimensions are significantly smaller than the axial dimension; material properties only vary along the length; the body force and the body couple are independent of the $y$ axis; loading condition does not lead to twisting; displacement field does not vary appreciably along height.

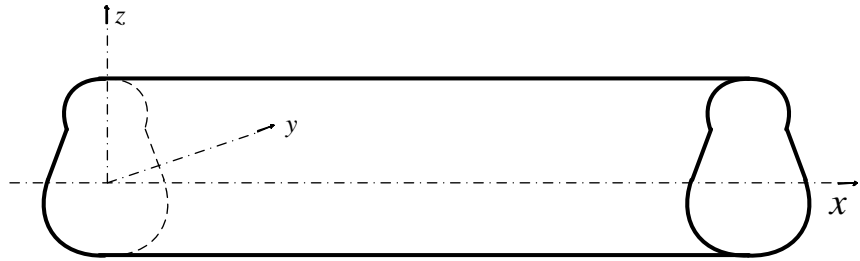

Figure 2. A schematic representation of the beam geometry

*4.2 Approximations to displacement and micro-rotation fields*

For a Timoshenko type beam, the three dimensional displacement and micro-rotation fields are approximated as follows.

$$u_x(x,y,z,t) = u(x,t) + z\psi(x,t) \ , \ u_y(x,y,z,t) = 0 \ , \ u_z(x,y,z,t) = w(x,t), \tag{40}$$

$$\theta_x(x,y,z,t) = 0 \ , \ \theta_y(x,y,z,t) = \theta(x,t) \ , \ \theta_z(x,y,z,t) = 0 \ , \tag{41}$$

where $u_x, u_y, u_z$ and $\theta_x, \theta_y, \theta_z$ are the Cartesian components of the displacement and micro-rotation fields. The approximations above allow expressing the three dimensional fields via several one dimensional descriptors, namely axial displacement $u$, macro-rotation $\psi$, transverse displacement $w$ and micro-rotation $\theta$ about the $y$ axis.

*4.3 1D kinematic states*

The necessary one dimensional kinematic states for this beam model are listed below.

Relative axial displacement state $\underline{u}[x]\langle x' - x \rangle = u(x') - u(x)$

Relative transverse displacement state $\underline{w}[x]\langle x' - x \rangle = w(x') - w(x)$

Relative macro-rotation state $\underline{\psi}[x]\langle x' - x \rangle = \psi(x') - \psi(x)$

Relative micro-rotation state $\underline{\theta}[x]\langle x' - x \rangle = \theta(x') - \theta(x)$

Average macro-rotation state $\underline{\hat{\psi}}[x]\langle x' - x \rangle = \frac{1}{2}(\psi(x') + \psi(x))$

Average micro-rotation state $\underline{\hat{\theta}}[x]\langle x' - x \rangle = \frac{1}{2}(\theta(x') + \theta(x))$

These states will be utilized to define non-local one dimensional strain-like quantities and subsequently the constitutive relations for several one dimensional generalized force states (to be introduced in the next section) will also be given in terms of these kinematic states.

*4.4 Equations of motion*

For the beam model at hand, the necessary equations of motion is obtainable by integrating the three dimensional equations of motions (Eqs. (2) and (14)) over the cross-section and introducing several one dimensional generalized force states. These equations have the following form.

$$m\ddot{u}(x,t) = \int_{x-\delta}^{x+\delta} \left( \underline{\mathbf{N}}_x[x]\langle x' - x \rangle - \underline{\mathbf{N}}_x[x']\langle x - x' \rangle \right) dx' + \mathbf{B}_x(x,t) \tag{42}$$

$$m\ddot{w}(x,t) = \int_{x-\delta}^{x+\delta} \left( \underline{\mathbf{N}}_z[x]\langle x' - x \rangle - \underline{\mathbf{N}}_z[x']\langle x - x' \rangle \right) dx' + \mathbf{B}_z(x,t) \tag{43}$$

$$\tilde{J}\ddot{\theta}(x,t) = \int_{x-\delta}^{x+\delta} \left( \underline{\mathcal{M}}_y[x]\langle x' - x \rangle - \underline{\mathcal{M}}_y[x']\langle x - x' \rangle \right) dx' + \frac{1}{2} \int_{x-\delta}^{x+\delta} \left( \underline{\mathbf{N}}_z[x]\langle x' - x \rangle - \underline{\mathbf{N}}_z[x']\langle x - x' \rangle \right)(x - x') dx'$$

$$+ \frac{1}{2} \int_{x-\delta}^{x+\delta} \left( \underline{\mathbf{M}}_s[x]\langle x' - x \rangle + \underline{\mathbf{M}}_s[x']\langle x - x' \rangle \right) dx' + \mathbf{L}_y(x,t) \tag{44}$$

The following equation is required in addition to Eqs. (42)-(44). It is obtained by multiplying the first of equation (2), i.e. the equation for the motion in the $x$ direction, with $z$ and then integrating over the cross-sectional area.

$$I\rho\ddot{\psi}(x,t) = \int_{x-\delta}^{x+\delta}\left(\underline{\mathbf{M}}[x]\langle x'-x\rangle - \underline{\mathbf{M}}[x']\langle x-x'\rangle\right)dx' - \frac{1}{2}\int_{x-\delta}^{x+\delta}\left(\underline{\mathbf{M}}_s[x]\langle x'-x\rangle + \underline{\mathbf{M}}_s[x']\langle x-x'\rangle\right)dx' + \mathrm{L}_z(x,t) \tag{45}$$

Definitions of the force states and moment states introduced in Eqs. (42)-(45) are listed below.

$$\underline{\mathbf{N}}_x[x]\langle x'-x\rangle = \int_A\int_A\left(\underline{\mathbf{T}}[\mathbf{X}]\langle x'-x\rangle\right)_x dA'dA \ , \ \underline{\mathbf{N}}_z[x]\langle x'-x\rangle = \int_A\int_A\left(\underline{\mathbf{T}}[\mathbf{X}]\langle x'-x\rangle\right)_z dA'dA \ ,$$

$$\underline{\mathcal{M}}_y[x]\langle x'-x\rangle = \int_A\int_A\left(\underline{\boldsymbol{\mu}}[\mathbf{X}]\langle x'-x\rangle\right)_y dA'dA \ , \ \underline{\mathbf{M}}_s[x]\langle x'-x\rangle = \int_A\int_A\left(\underline{\mathbf{T}}[\mathbf{X}]\langle x'-x\rangle\right)_x (z'-z) dA'dA \ ,$$

$$\underline{\mathbf{M}}[x]\langle x'-x\rangle = \frac{1}{2}\int_A\int_A\left(\underline{\mathbf{T}}[\mathbf{X}]\langle x'-x\rangle\right)_x (z'+z) dA'dA \ .$$

Also the integrated body forces and couples that appear in the beam equations of motion are derived as

$$\mathrm{B}_x(x,t) = \int_A \mathrm{b}_x dA \ , \ \mathrm{B}_z(x,t) = \int_A \mathrm{b}_z dA \ , \ \mathrm{L}_z(x,t) = \int_A z\,\mathrm{b}_x dA \ , \ \mathrm{L}_y(x,t) = \int_A \mathrm{l}_y dA \ .$$

Further, $m = \int_A \rho\, dA$, $\tilde{J} = \int_A J\, dA$. The area element $dA = dydz$ and $dA' = dy'dz'$. $\delta$ is the maximum (radial) distance over which the long range interaction is considered.

*4.5 Constitutive equations*

To find the constitutive relations for these one dimensional non-local force and moment states, we define the internal energy per unit length ($e_1$) of the beam. Eq. (30), using approximate deformation fields (see Section 4.2), is integrated over the cross-sectional area leading to the following expression.

$$\dot{e}_1 = \int_{x-\delta}^{x+\delta}\underline{\mathbf{N}}_x\langle x'-x\rangle\underline{\dot{u}}\langle x'-x\rangle dA' + \int_{x-\delta}^{x+\delta}\underline{\mathbf{N}}_z\langle x'-x\rangle\left(\underline{\dot{w}}\langle x'-x\rangle + \underline{\dot{\theta}}\langle x'-x\rangle(x'-x)\right)dA'$$

$$+ \int_{x-\delta}^{x+\delta}\underline{\mathcal{M}}_y\langle x'-x\rangle\underline{\dot{\theta}}\langle x'-x\rangle dA' + \int_{x-\delta}^{x+\delta}\underline{\mathbf{M}}\langle x'-x\rangle\underline{\dot{\psi}}\langle x'-x\rangle dA'$$

$$+ \int_{x-\delta}^{x+\delta}\underline{\mathbf{M}}_s\langle x'-x\rangle\left(\underline{\dot{\psi}}\langle x'-x\rangle - \underline{\dot{\theta}}\langle x'-x\rangle\right)dA' \tag{46}$$

Using a similar definition for inner product of states in one dimension and introducing the required composite states, the equation above can be written in the following compact form.

$$\dot{e}_1 = \underline{\mathbf{N}}_x\bullet\underline{\dot{u}} + \underline{\mathbf{N}}_z\bullet\underline{\dot{w}}_{\theta x} + \underline{\mathcal{M}}_y\bullet\underline{\dot{\theta}} + \underline{\mathbf{M}}\bullet\underline{\dot{\psi}} + \underline{\mathbf{M}}_s\bullet\underline{\dot{\psi}}_{\hat{\theta}} \ , \tag{47}$$

where $\underline{w}_{\theta x}\langle x'-x\rangle = \underline{w}\langle x'-x\rangle + \underline{\hat{\theta}}\langle x'-x\rangle(x'-x)$ and $\underline{\psi}_{\hat{\theta}}\langle x'-x\rangle = \underline{\hat{\psi}}\langle x'-x\rangle - \underline{\hat{\theta}}\langle x'-x\rangle$ .

Assuming the energy density functional $e_1$ to depend on the states $\underline{u}$, $\underline{w}_{\theta x}, \underline{\theta}$, $\underline{\psi}$ and $\underline{\psi}_{\hat{\theta}}$, similar constitutive relations as in Eq. (33) may be obtained as

$$\underline{N}_x = e_{1\underline{u}} \ , \ \underline{N}_z = e_{1\underline{w}_{\theta x}} \ , \ \underline{\mathcal{M}}_y = e_{1\underline{\theta}} \ , \ \underline{M} = e_{1\underline{\psi}} \text{ and } \underline{M}_s = e_{1\underline{\psi}_\theta} \ . \tag{48}$$

*4.5.1 Constitutive correspondence*

A similar strategy as described in Section 3.3.1 is adopted here to establish the constitutive correspondence in one dimension. In the classical micropolar description of a beam (Ramezani et al., 2009), an expression for strain energy rate per unit length in terms of one dimensional forces and moments (or bending moment) and their conjugate strain-like quantities can be written as follows.

$$\dot{w}_1 = AN_{xx}\dot{\gamma}_u + AN_{xz}\dot{\gamma}_{w\theta} + AN_{zx}\dot{\gamma}_{\psi\theta} + A\mathcal{M}_{xy}\dot{\kappa}_\theta + M\dot{\kappa}_\psi \tag{49}$$

where $\gamma_u = \dfrac{\partial u}{\partial x}$ , $\gamma_{w\theta} = \dfrac{\partial w}{\partial x} + \theta$ , $\gamma_{\psi\theta} = \psi - \theta$ , $\kappa_\theta = \dfrac{\partial \theta}{\partial x}$ and $\kappa_\psi = \dfrac{\partial \psi}{\partial x}$ .

Definitions of the one dimensional forces and moments used in (49) along with the constitutive relations (see Eq. (9)) are given below.

$$N_{xx} = \frac{1}{A}\int_A \sigma_{xx}dA = E\gamma_u \ , \ N_{xz} = \frac{1}{A}\int_A \sigma_{xz}dA = (\tilde{\mu}+\eta)\gamma_{w\theta} + \tilde{\mu}\gamma_{\psi\theta} \ , \ N_{zx} = \frac{1}{A}\int_A \sigma_{zx}dA = (\tilde{\mu}+\eta)\gamma_{\psi\theta} + \tilde{\mu}\gamma_{w\theta},$$

$$\mathcal{M}_{xy} = \frac{1}{A}\int_A \mu_{xy}dA = \beta\kappa_\theta \text{ and } M = \frac{1}{A}\int_A z\sigma_{xx}dA = EI\kappa_\psi \ . \tag{50}$$

We introduce the following one dimensional, non-local, strain-like quantities that are useful in establishing the constitutive correspondence.

$$\overline{\gamma}_u(\underline{u}) = \left[\int_{x-\delta}^{x+\delta} \omega_1(|x'-x|)\underline{u}\langle x'-x\rangle(x'-x)dx'\right]\overline{K}_1^{-1}$$

$$\overline{\gamma}_{w\theta}(\underline{w}_{\theta x}) = \left[\int_{x-\delta}^{x+\delta} \omega_1(|x'-x|)\underline{w}_{\theta x}\langle x'-x\rangle(x'-x)dx'\right]\overline{K}_1^{-1}$$

$$\overline{\gamma}_{\psi\theta}(\underline{\psi}_\theta) = \left[\int_{x-\delta}^{x+\delta} \omega_1(|x'-x|)\underline{\psi}_\theta\langle x'-x\rangle(x'-x)^2 dx'\right]\overline{K}_1^{-1}$$

$$\overline{\kappa}_\theta(\underline{\theta}) = \left[\int_{x-\delta}^{x+\delta} \omega_1(|x'-x|)\underline{\theta}\langle x'-x\rangle(x'-x)dx'\right]\overline{K}_1^{-1}$$

$$\overline{\kappa}_\psi(\underline{\psi}) = \left[\int_{x-\delta}^{x+\delta} \omega_1(|x'-x|)\underline{\psi}\langle x'-x\rangle(x'-x)dx'\right]\overline{K}_1^{-1} \tag{51}$$

Here $\omega_1(|x'-x|) > 0$ for $|x'-x| \leq \delta$ and $\omega_1(|x'-x|) = 0$ otherwise. The one dimensional shape tensor is given by $\overline{K}_1 = \int_{x-\delta}^{x+\delta} \omega_1(|x'-x|)(x'-x)^2 dx'$ .

$\overline{\gamma}_u$, $\overline{\gamma}_{w\theta}$, $\overline{\gamma}_{\psi\theta}$, $\overline{\kappa}_\theta$ and $\overline{\kappa}_\psi$ are respectively the non-local approximations to $\gamma_u$, $\gamma_{w\theta}$, $\gamma_{\psi\theta}$, $\kappa_\theta$ and $\kappa_\psi$. Replacing the local strain terms in Eq. (49) by their non-local counterparts and using the fact that $\dot{e}_1 = \dot{w}_1$, the following relations are obtained using Eq. (46).

$$\underline{\mathbf{N}}_x \langle x'-x \rangle = \omega_1(|x'-x|)(AE\overline{\gamma}_u)\overline{\mathbf{K}}_1^{-1}(x'-x)$$

$$\underline{\mathbf{N}}_z \langle x'-x \rangle = \omega_1(|x'-x|)A\left((\tilde{\mu}+\eta)\overline{\gamma}_{\psi\theta} + \tilde{\mu}\overline{\gamma}_{w\theta}\right)\overline{\mathbf{K}}_1^{-1}(x'-x)$$

$$\underline{\mathcal{M}}_y \langle x'-x \rangle = \omega_1(|x'-x|)(A\beta\overline{\kappa}_\theta)\overline{\mathbf{K}}_1^{-1}(x'-x)$$

$$\underline{\mathbf{M}} \langle x'-x \rangle = \omega_1(|x'-x|)(EI\overline{\kappa}_\psi)\overline{\mathbf{K}}_1^{-1}(x'-x)$$

$$\underline{\mathbf{M}}_s \langle x'-x \rangle = \omega_1(|x'-x|)A\left((\tilde{\mu}+\eta)\overline{\gamma}_{w\theta} + \tilde{\mu}\overline{\gamma}_{\psi\theta}\right)\overline{\mathbf{K}}_1^{-1}(x'-x)^2 \tag{52}$$

Eqs. (42)-(45) and the constitutive relations in (52) completely describe the micropolar Timoshenko type PD beam.

## 5. Non-polar PD beam

For the state based non-polar PD as described in Section 2.1, a similar beam model is derived. The beam geometry, different relevant assumptions, the approximation of the displacement field and the necessary states are similar to those noted in Section 4. Dynamics of this non-polar 1D model is captured through the following equations of motion.

$$m\ddot{u}(x,t) = \int_{x-\delta}^{x+\delta} \left(\underline{\mathbf{N}}_x[x]\langle x'-x \rangle - \underline{\mathbf{N}}_x[x']\langle x-x' \rangle\right)dx' + \mathbf{B}_x(x,t) \tag{53}$$

$$m\ddot{w}(x,t) = \int_{x-\delta}^{x+\delta} \left(\underline{\mathbf{N}}_z[x]\langle x'-x \rangle - \underline{\mathbf{N}}_z[x']\langle x-x' \rangle\right)dx' + \mathbf{B}_z(x,t) \tag{54}$$

$$I\rho\ddot{\psi}(x,t) = \int_{x-\delta}^{x+\delta} \left(\underline{\mathbf{M}}[x]\langle x'-x \rangle - \underline{\mathbf{M}}[x']\langle x-x' \rangle\right)dx' + \frac{1}{2}\int_{x-\delta}^{x+\delta} \left(\underline{\mathbf{N}}_z[x]\langle x'-x \rangle - \underline{\mathbf{N}}_z[x']\langle x-x' \rangle\right)(x-x')dx'$$
$$+ \mathbf{L}_z(x,t) \tag{55}$$

The above equations are derived following the same steps as outlined in Section 4. For the derivation of Eq. (55), constitutive restrictions for the non-polar material given by Eq. (3) are imposed in addition. The force states, body force and couple have the same definitions given in the previous section.

The required constitutive equations are

$$\underline{\mathbf{N}}_x \langle x'-x \rangle = \omega_1(|x'-x|)(AE\overline{\gamma}_u)\overline{\mathbf{K}}_1^{-1}(x'-x), \quad \underline{\mathbf{N}}_z \langle x'-x \rangle = \omega_1(|x'-x|)(AG\overline{\gamma}_{w\psi})\overline{\mathbf{K}}_1^{-1}(x'-x),$$
$$\underline{\mathbf{M}} \langle x'-x \rangle = \omega_1(|x'-x|)(EI\overline{\kappa}_\psi)\overline{\mathbf{K}}_1^{-1}(x'-x), \tag{56}$$

where the non-local strains, except $\overline{\gamma}_{w\psi}$, are defined in Eq. (51). Definition of $\overline{\gamma}_{w\psi}$ is as follows.

$$\overline{\gamma}_{w\psi}(\underline{\mathbf{w}}_{\tilde{\psi}x}) = \left[\int_{x-\delta}^{x+\delta} \omega_1(|x'-x|)\underline{\mathbf{w}}_{\tilde{\psi}x}\langle x'-x \rangle(x'-x)dx'\right]\overline{\mathbf{K}}_1^{-1}$$

with $\underline{\mathbf{w}}_{\tilde{\psi}x}\langle x'-x \rangle = \underline{\mathbf{w}}\langle x'-x \rangle + \underline{\hat{\psi}}\langle x'-x \rangle(x'-x)$.

## 6. Numerical Illustrations

In this section, two illustrative examples, viz. the bending of a thin beam and the in-plane tensile response of a plate with a circular hole, are considered to demonstrate the effects of micropolarity as incorporated in the state based PD theory. While the proposed one-dimensional model (see Sections 4 and 5) is employed for the bending analysis of a beam, a somewhat straightforward 2D plane-stress type reduction of the 3D theory, the details of which are given in Section 3, is adopted for the analysis of a plate with a hole.

For the purpose of numerical implementation, body (1D or 2D) is discretized into a finite number of nodes in its reference configuration. Each node is associated with finite length in case of 1D body and finite area in the 2D case. These nodes are typically termed as PD particles. All the governing equations are then written for these PD particles along with a Riemann-sum type approximation (summed over these PD particles) for the integral quantities in these equations (Eqs.(2), (14), (42)-(45), (53)-(55)). This discretization results in a set of algebraic equations with displacement and microrotation at the different PD particles as unknowns, which are solved for after proper imposition of boundary conditions. Details of numerical implementation of a typical PD algorithm may be found elsewhere, e.g. Breitenfeld et al. (2014).

*6.1. One-dimensional beam*

Deformation characteristics of nano- or micro-beams are strongly affected by the internal length scale parameters of the constituent materials. Such beams generally show stiffer response than what is predicted by classical (local) theories. The proposed one dimensional micropolar PD beam model is expected to capture such stiffer response correctly owing to the extra information it carries on the material length scale parameters. One may contrast this with the non-polar PD, which should fail to predict such response despite carrying some length scale information that, whilst being inherent in a PD formulation, may still be physically irrelevant.

For the numerical implementation, the geometric and material properties considered for the beam model are:

$L = 1 m$, $h = 0.05$ m, $b = 0.05$ m, $E = 20$ GPa, $\nu = 0.3$, $\eta = G/20$, $\beta = G/5000$.

where $L$, $h$ and $b$ are the length, thickness and width of beam respectively. The beam is subject to the following loading conditions.

$B_x = 0$, $B_z = 10^3$ N/m and $L_y = 0$ N.m/m.

Please note that while these parameters may not be physically representative, they are adequate for the purpose of demonstration of our technique.

The beam is analyzed for three different boundary conditions, namely fixed-free (a cantilever beam), fixed-fixed (a fixed beam) and pinned-pinned (a simply supported beam), using both non-polar and micropolar PD models. As a standard practice in PD, the boundary conditions are applied, in each case, over a patch of nodes considered beyond the actual length of the beam. With $l = \sqrt{\beta/(2G)}$ as the definition for the length scale parameter, it may be observed that the beam thickness considered is of the same order as $l = 0.01$ and hence it is anticipated that the length scale parameter would considerably influence the deformation characteristic of the beam. As reported in Figures 3 and 4, the non-polar PD solutions are seen to be insensitive to the material length scale parameter. Exact solutions considering the Euler-Bernoulli model are also reported in Figures 3 and 4. Low thickness to length ratio of the beams allows the Euler-Bernoulli type model to be applicable in this case. Both Euler-Bernoulli and non-polar Timoshenko-type PD models yield comparable results and fail to reflect the anticipated stiffening of the beams at that length scale. The proposed micropolar PD model, on the other hand, successfully brings out the expected length scale effects in the transverse deflection and rotation profiles of the beam (see Figures 3 and 4). Figure 5 shows the computed micro-rotations for each of the three beams.

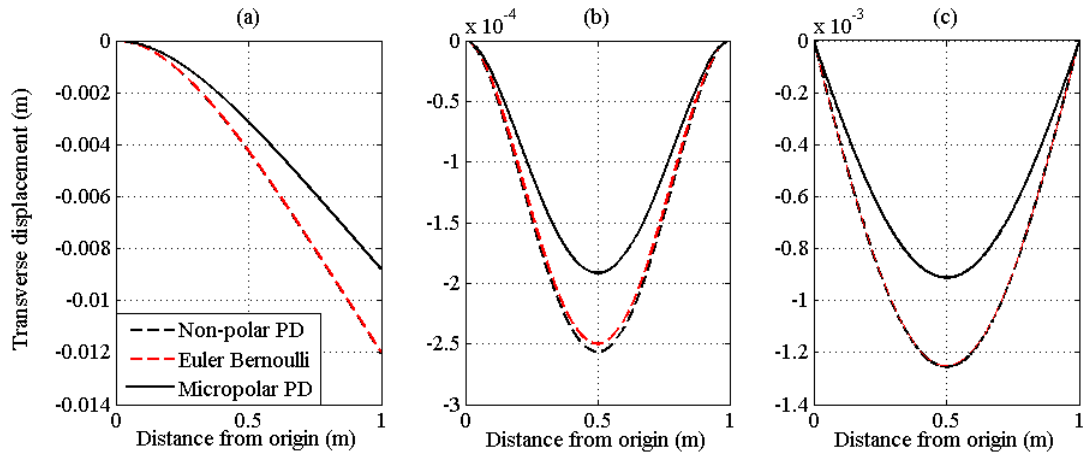

Figure 3. Transverse displacement of one dimensional beam:

(a) Cantilever Beam (b) Fixed Beam and (c) Simply Supported Beam

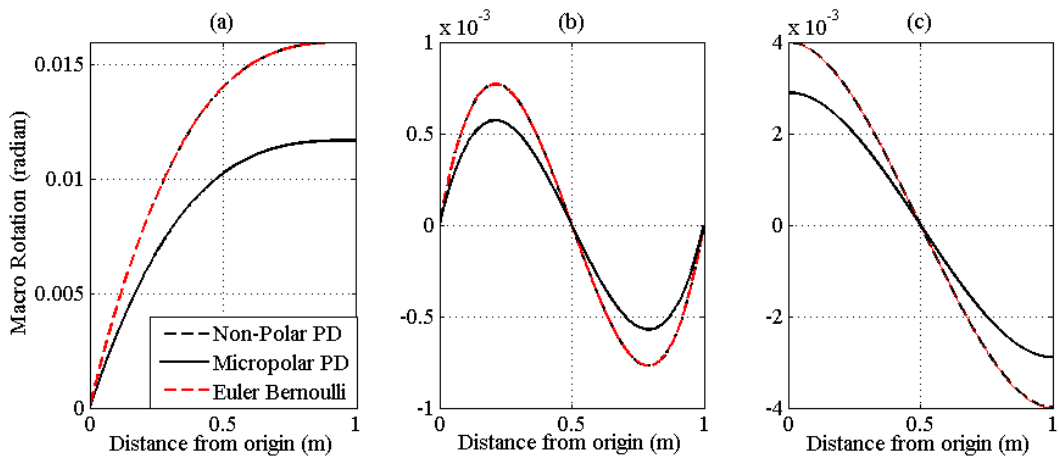

Figure 4. Macro rotation of one dimensional beam:

(a) Cantilever Beam (b) Fixed Beam and (c) Simply Supported Beam

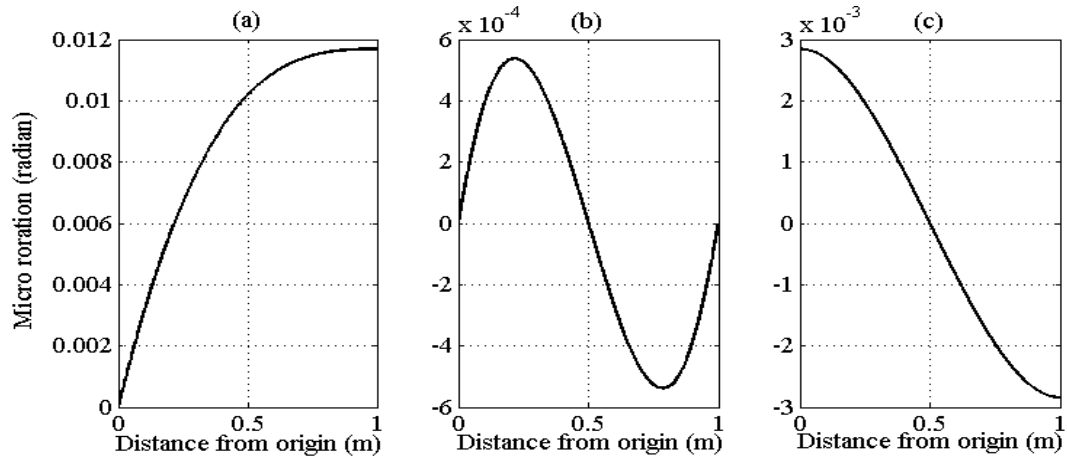

Figure 5. Micro rotation of one dimensional beam:

(a) Cantilever Beam (b) Fixed Beam and (c) Simply Supported Beam

An additional numerical parametric study is done for the cantilever beam to demonstrate the effect of the length scale on its transverse tip deflection. Keeping the length scale parameter fixed at 0.01 m, as the thickness is reduced, the deflection tends to blow up when the non-polar PD model is employed. However the proposed micropolar PD model remedies the situation by obtaining a deformation profile whose magnitude is smaller by many orders (Figure 6a) and thus remains consistent with the physically expected scenario. Figure 6(b) shows the result of another study where the cantilever beam thickness is kept fixed at 0.05 m and the tip deflection is reported for different length scale values. While the non-polar PD model again shows up length scale insensitive tip deflection, the micropolar PD deflection is clearly affected by the varying length scale. As the length scale value is reduced from 0.05 m to zero, the length scale effect diminishes and both the micropolar PD and non-polar PD schemes lead to the same solution.

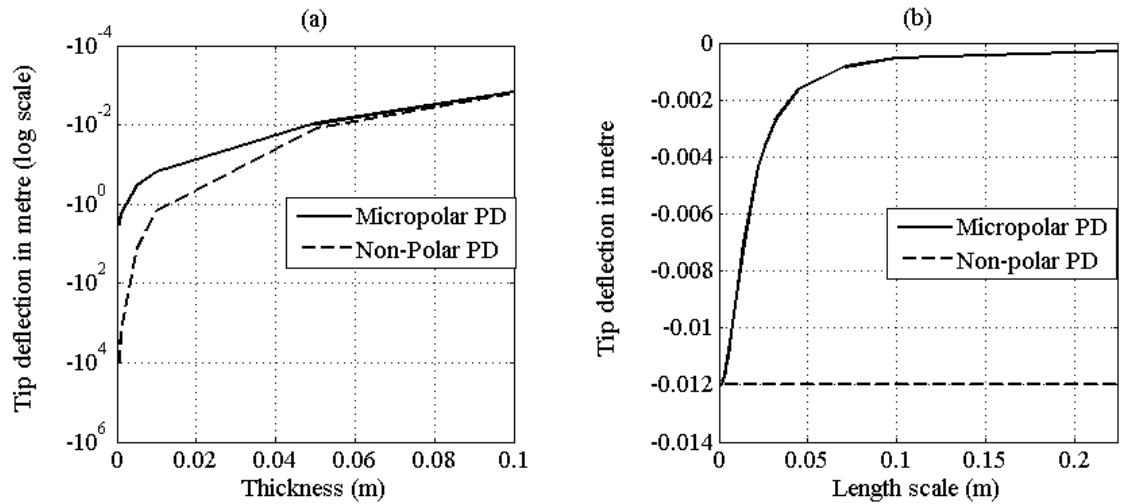

Figure 6. Tip deflection of cantilever beam

(a) Thickness vs. tip deflection for constant length scale (0.01 m) (b) length scale vs. tip deflection for constant thickness (0.05 m)

*6.2. Elastic plate with a hole*

The static in-plane tensile response of a plate with a central circular hole is known to display a strong influence of the material length scale parameter as the hole radius becomes equal or less than the length scale value (Mindlin, 1963). In such a case, the stress concentration near the hole is significantly lesser than what is predicted by the classical theory. Therefore, in the context of 2D elastostatic problems, this example is chosen here to verify the effectiveness of the proposed micropolar PD theory over the standard non-polar one.

To reduce the computational overhead, a plate of very small dimensions (length 0.02 m, width 0.005 m and thickness 0.001 m with a central circular hole of radius $a = 0.00125$ m) is considered here for illustrative purposes. The plate is subjected to a uniaxial uniform tension $p_0 = 10^6$ N/m as shown in Figure 7. Plane-stress constitutive relations are used to solve this problem using both non-polar PD and proposed micropolar PD models. The plate material is assumed to have the following properties:

$E = 100 \text{GPa}$, $v = 0.3$, $\eta = G/1.5$, $\beta = G/160000$.

In order to make the length scale effect conspicuous, the hole radius is chosen less than the length scale parameter value (the latter being 0.0018 m) as defined in Section 6.1.

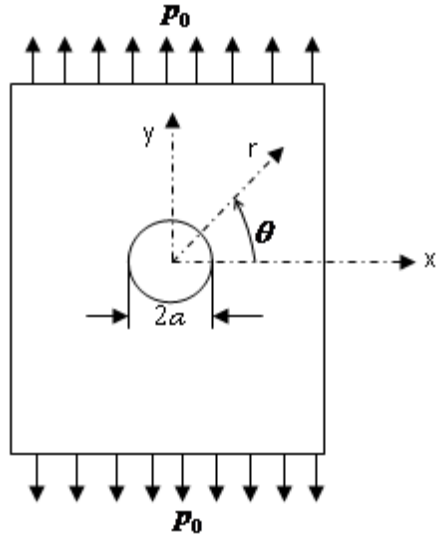

Figure 7. Plate with a hole subjected to uniform tension

Although the force and moment states are obtained by solving the micropolar PD model, we have chosen to report the components of the equivalent nonlocal Cauchy-type stress in order to display the results in more familiar forms. Displacement and micro-rotation vectors, obtained by solving the PD model, are used to compute the non-local strain and wryness using the definitions given in (36). Constitutive relations given in (9) are appropriately reduced for the plane stress case and evaluated using the non-local strain and wryness. Non-local stress and couple stress components thus computed are reported in some of the following figures. A similar treatment is adopted to represent the results through the non-polar PD model too. Figure 8 shows the normal stress ($\sigma_{yy}$) distributions via both the PD models.

It may be observed that the incorporation of the length scale parameter in the micropolar PD results in lower normal stresses near the hole than those through the non-polar PD model by a factor of nearly 1.3. While Figure 9(a) shows the shear stress components obtained in the non-polar PD case to be symmetric, the asymmetric nature of the shear stress components ($\sigma_{xy} \neq \sigma_{yx}$) based on the micropolar PD theory is evident in Figures 9(b, c). Distributions of non-zero couple stresses, especially near the hole, in the micropolar PD model are shown in Figure 10. Such stresses are ignored in the non-polar PD variant. Essentially the consideration of couple stresses in the micropolar PD theory entails an inherent asymmetry of the stress tensor and an incorporation of the intrinsic material length parameter is responsible for reducing the normal stresses in the vicinity of the hole.

In order to provide an explicit comparison of the normal stresses ($\sigma_{yy}$) obtained from both the models, the normalized stress distribution ($\sigma_{yy}/p_0$) along the radial line $\theta = 0^0$ is reported in Figure 11. This figure clearly shows that the stress concentration factor (defined as the normalized stress at the hole edge i.e. at $r = a$ and $\theta = 0°$) is predicted higher in non-polar PD case, with a much lesser concentration being obtained when the micropolar PD model is solved. Effect of the length scale in reducing the stress concentration is thus numerically validated through the proposed micropolar PD model.

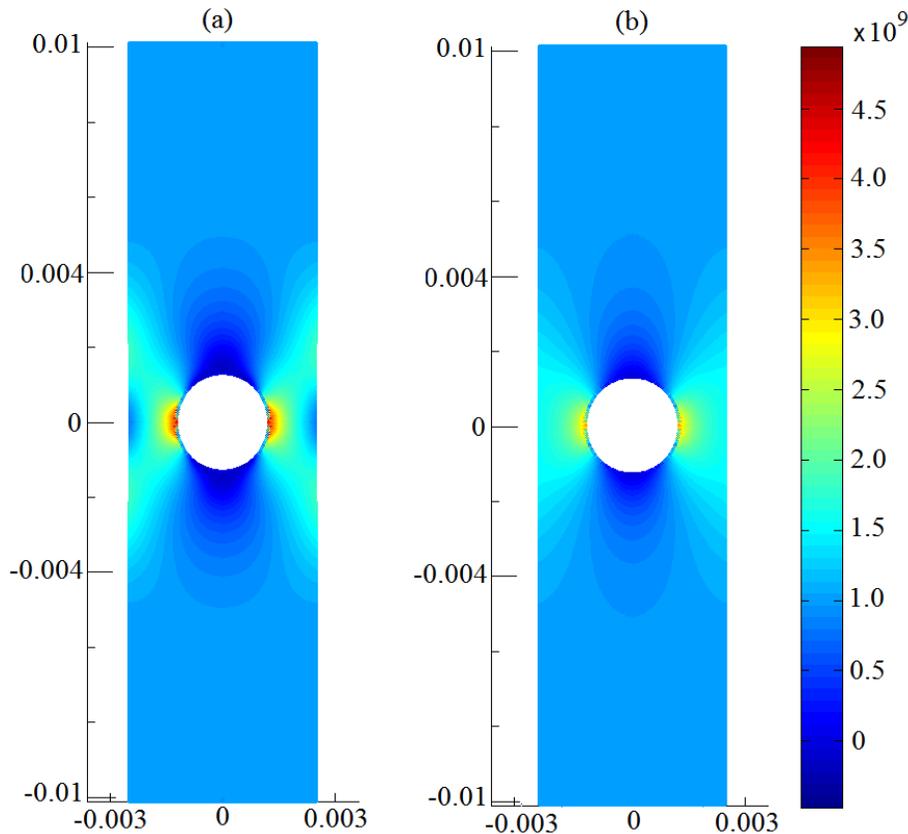

Figure.8. Stress ($\sigma_{yy}$) distribution: (a) Non-polar PD (b) Micropolar PD

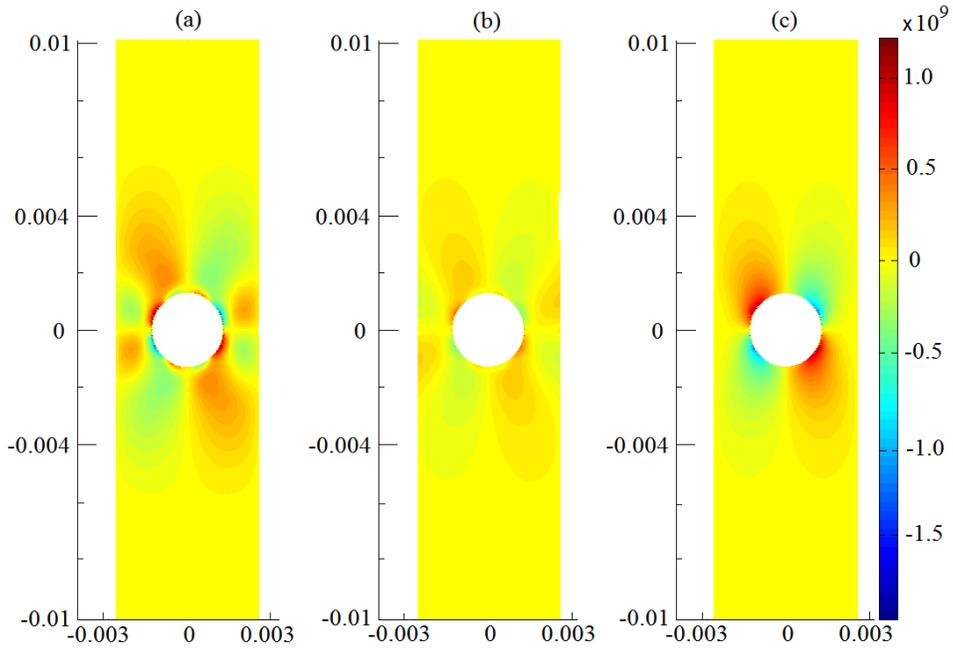

Figure.9. Shear stresses: (a) $\sigma_{xy} = \sigma_{yx}$ (Non-polar PD) (b) $\sigma_{xy}$ (Micropolar PD) (c) $\sigma_{yx}$ (Micropolar PD)

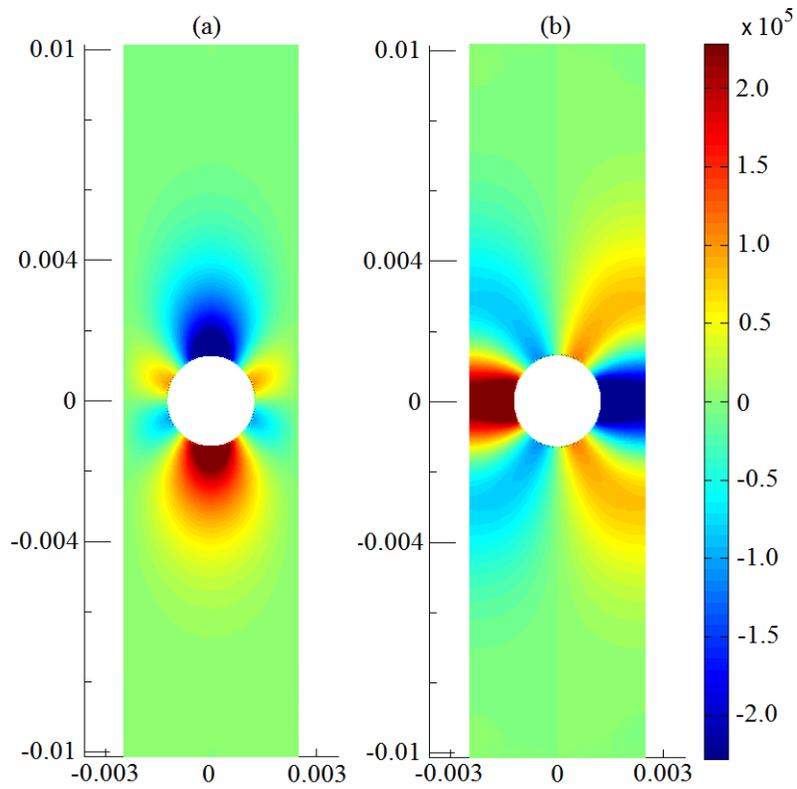

Figure.10. Couple stresses: (a) $\mu_{zx}$ (Micropolar PD ) (b) $\mu_{zy}$ (Micropolar PD)

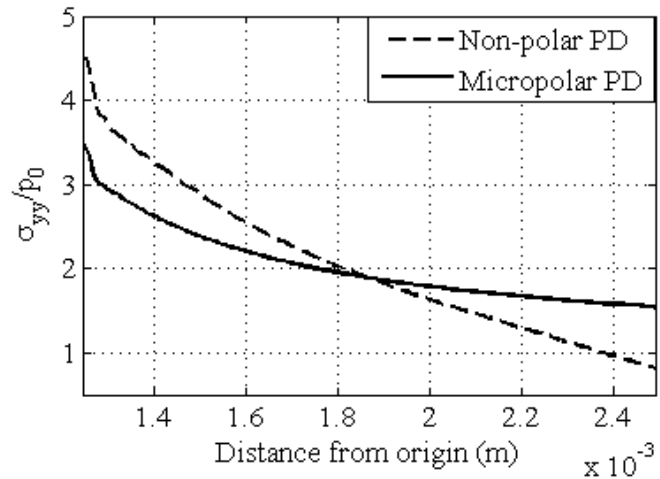

Figure.11. Normalized stress ($\sigma_{yy}/p_0$) distribution along $\theta = 0°$

## 7. Conclusions

A three dimensional, state-based micropolar PD model has been developed, incorporating physically consistent nonlocal particle interactions through the use of an additional length scale parameter. This length scale parameter is in addition to the horizon radii present in the standard PD theory. The material model in this work has been restricted to the micropolar isotropic elastic case. The incorporation of micropolar effects in the PD model equips it to better predict the deformation of continua where the length scale effect is significant, e.g. very thin beams. It is also expected to be useful to model other, similar continua like nano-beams and nano-sheets. In addition, homogenized one dimensional micropolar as well as non-polar PD beam models have been derived. In lieu of the full-blown 3D continuum, the use of dimensionally reduced models such as the micropolar PD beam involve significantly lower computational overhead. The non-polar variant of the one dimensional PD model is shown to be material length scale insensitive and therefore predicts unphysically large deformation. This is remedied by developing a micropolar PD model that accounts for length scale effects, thus restoring the physicality of the response. A similar observation holds even for planar elastostatic problems. An application of the proposed micropolar PD model for dynamical systems and its extensions incorporating plasticity and damage will be considered elsewhere. Indeed, it is with such problems as damage propagation, which violate material continuity and whose evolution necessarily demands length scales, that the full spectrum of advantages of the proposed micropolar PD scheme will be adequately exploited.